\documentclass[aip, jap, 10pt, twocolumn, reprint, superscriptaddress, amsmath, amssymb]{revtex4-2} 
\usepackage{graphicx}
\usepackage{dcolumn}
\usepackage{bm}

\usepackage{float}
\usepackage{hyperref}

\usepackage[usenames]{xcolor}

\usepackage{multirow}
\usepackage{threeparttable}

\usepackage{changes}
\definechangesauthor[name={Sven C. Liebscher},color=blue]{SL}
\definechangesauthor[name={Lars C. Bannow},color=green]{LB}

\begin{document}
\begin{center}
\Huge
 The following article has been submitted to Journal of Applied Physics. After it is published, it will be found at \url{http://jap.aip.org}

\end{center}
\thispagestyle{empty}
\clearpage

\newfloat{scheme}{hhtbp}{lof}
\floatname{scheme}{\small SCH.~}

\preprint{APS/123-QED}

\title{Extension of the LDA-1/2 method to the material class of bismuth containing III-V semiconductors}

\author{Sven C. Liebscher}
\author{Lars C. Bannow}%
\affiliation{ 
Fachbereich Physik, Philipps-Universit\"at Marburg, Renthof 5, 35032 Marburg, Germany
}%

\author{J{\"o}rg Hader}
\author{Jerome V. Moloney}
\affiliation{%
Wyant College of Optical Sciences, University of Arizona, Tucson, Arizona 85721, USA
}%

\author{Stephan W. Koch}
\affiliation{ 
Fachbereich Physik, Philipps-Universit\"at Marburg, Renthof 5, 35032 Marburg, Germany
}%
\affiliation{ 
Material Sciences Center, Philipps-Universit\"at Marburg, Hans-Meerwein-Stra\ss e 6, 35032 Marburg, Germany
}%
%
%

\date{\today}%
\setcounter{page}{1}
\begin{abstract}
The LDA-1/2 method is employed in density functional theory calculations for the electronic structure of III-V dilute bismide systems. For the representative example of Ga(SbBi) with Bi concentrations below $10 \%$, it is shown that this method works very efficiently, especially due to its reasonably low demand on computer memory. The resulting bandstructure and wavefunctions are used to compute the interaction matrix elements that serve as input to microscopic calculations of the optical properties and intrinsic losses relevant for optoelectronic applications of dilute bismides.


\end{abstract}


\maketitle


\section{Introduction}
In the semiconductor research community exists an ongoing interest in III-V materials that contain a dilute amount of bismuth (Bi) \cite{Wang2017}. These so called bismides are promising candidates for opto-electronic applications as light-emitting diodes \cite{Hossain2012}, semiconductor lasers \cite{Marko2016}, or solar cells \cite{Richards2017}.
As can be seen from the material combinations listed in Ref.~\onlinecite{Wang2017}, bismuth can nowadays be incorporated into almost every III-V material. Since this allows for many different material combinations, a predictive approach for their electronic and optical properties is desirable. 

For the example of In(AsBi), a scheme that combines density functional theory (DFT) and a microscopic many-body methodology has been applied and validated by detailed theory-to-experiment comparisons.\cite{Hader2018, Hader2018b} Unfortunately, however, the DFT modeling of dilute III-V material systems requires the use of very large supercells which are needed to approximate a random distribution of the relatively low concentration of the incorporated atoms.  Previously, it has been shown that reliable bandgap calculations for dilute nitrides based on DFT require supercells with up to $432$ atoms \cite{Rosenow2018}. Since commonly used functionals such as the local density approximation (LDA) \cite{Ceperley1980} or the Perdew-Burke-Ernzerhof generalized gradient approximation \cite{Perdew1996} severely underestimate the bandgap of semiconductors, the metaGGA TB09 \cite{Tran2009} had to be applied in Refs.~\onlinecite{Rosenow2018, Hader2018, Hader2018b}. In addition to a dependence on the electron density and its gradient, the TB09 functional also depends on the kinetic energy density. Therefore, DFT calculations utilizing this functional are in need of significantly more computational memory when compared to the LDA, which only depends on the electron density. This can become a limiting factor when not only the supercells are large but also many $k$-points have to be included in the calculation. For instance, this is the case when DFT is used to calculate the bandstructure and $k$-dependent dipole matrix elements as input for optical calculations. 

In order to overcome the existing limitations, in the current study, the LDA-1/2 method\cite{Ferreira2008, Ferreira2011a} is utilized which requires only as much memory as the LDA. The LDA-1/2 method is based on Slater's half-occupation technique \cite{Slater1972} and improves the Kohn-Sham bandgap towards the real bandgap of the system. Recently, the strength of the LDA-1/2 method when it comes to large supercells was emphasized \cite{Doumont2019}. The LDA-1/2 method depends on the variation of a cutoff radius until the bandgap of the system is maximized \cite{Ferreira2008}. However, since III-Bi materials are semi-metals \cite{bi_semi}, this procedure cannot be used to determine the cutoff radii for the Bi atoms in AlBi, GaBi and InBi. To overcome this problem, we use an alternative extrapolation method to obtain the cutoff radii for the Bi atoms which allows us to extend the LDA-1/2 method for application in dilute bismides. As a representative example, we calculate the electronic structure for Ga(SbBi) assuming different concentrations of bismuth below $10 \%$ and use the electronic structure results as input for the evaluation of the intrinsic optoelectronic properties.

\section{Computational Method\label{Sec:methods}}

\subsection{The LDA-1/2 method}

Details of the LDA-1/2 method, especially its theoretical formulation, are presented in Refs.~\onlinecite{Ferreira2008, Ferreira2011a}. Therefore, we restrict the discussion in this paper to the aspects important for the application of the LDA-1/2 method to dilute bismides. Ferreira \textit{et al.} \cite{Ferreira2008} introduce what they call a self-energy potential $V_S(\pmb{r})$ whose quantum-mechanical average is the self energy
\begin{equation}
 S_{\alpha} = \int \text{d}^3r\,n_{\alpha}(\pmb{r})V_S(\pmb{r})
\end{equation}
with state $\alpha$ and electron density $n_{\alpha}$. This self-energy potential $V_S(\pmb{r})$ is obtained for the specific atom by subtracting the Kohn-Sham effective potential of the half-ionized state from that of the neutral state. Subsequently, for calculations of solids, $V_S(\pmb{r})$ is added to the LDA pseudopotentials. However, as $V_S(\pmb{r}) \rightarrow 1/r$ for long ranges this would result in a divergent result. Therefore, $V_S(\pmb{r})$ is first multiplied by a spherical step function given as
\begin{equation}
 \Theta(r) = 
 \begin{cases}
  \left[1-\left(\frac{r}{r_\text{cut}}\right)^n\right]^3 & r \leq r_\text{cut} \\
  0 & r > r_\text{cut}
 \end{cases}
\end{equation}
where the usual choice is $n = 8$ \cite{Ferreira2008}. As shown by Xue \textit{et al.} \cite{Xue2018}, for $n = 8$ the tails of the step function are long such that larger values of $n$ are better suited for adding $V_S(\pmb{r})$ only in the region where the valence hole is located. Following Ref.~\onlinecite{Xue2018}, we therefore set $n = 20$ in our calculations. The precise cutoff radius $r_\text{cut}$ is determined by its variation until the bandgap of the solid in question is maximized. At this choice for the cutoff radius as much as possible of the region that is occupied by the valence band hole is covered by the sphere \cite{Xue2018}. Furthermore, following Refs.~\onlinecite{Ferreira2008, Ferreira2011a}, here the LDA-1/2 method is only applied to the anions as the valence band states usually originate from these.

\subsection{Supercell generation}
\label{sup}
DFT calculations were performed with the Vienna \textit{ab initio} Simulation Package (VASP 5.4.4)\cite{Kresse1993, Kresse1994, Kresse1996, Kresse1996a} which uses a plane-wave basis set and the projector-augmented wave method\cite{Blochl1994, Kresse1999}. In our evaluations,
a 370 eV cutoff energy of the plane-wave basis was used and
the criteria for electronic convergence and ionic relaxation were set to $10^{-7}\;\text{eV}$ and $10^{-6}\;\text{eV}\cdot \text{\AA}^{-1}$, respectively. 
For relaxations, the LDA\cite{Perdew1981} exchange-correlation potential was used and 
a $\Gamma$-centered Monkhorst-Pack-grid\cite{Monkhorst1976} of 8x8x8 k-points was implemented for primitve cell calculations.
After relaxation of the primitive cells of GaSb and GaBi, supercells of different geometries were created using the Alloy Theoretic Automated Toolkit (ATAT) \cite{vandeWalle2013}.
The supercells consist of 64 primitive cells of GaSb, arranged in different cuboidal geometries with up to 8 primitive cells in one direction. 
Since the $x$ and $y$-directions are equivalent, geometries with permutations of $x$ and $y$ can be omitted, leaving the geometries 1x8x8, 2x4x8, 2x8x4, 4x4x4, 4x8x2 and 8x8x1.
The $k$-point grid was reduced accordingly for the supercells.
Special quasirandom structures (SQS)\cite{Zunger1990} were generated to stastically determine an optimal approximation to the random placement of the Bi atoms in the supercell.
Up to 4 Bi atoms were incorporated into the supercells, yielding the concentrations $0\%, 1.5625\%, 3.125\%, 4.6875\%$ and $6.25\%$.

The relaxations of the supercells were constrained to the z-direction in order to simulate the effect of a substrate with fixed in-plane lattice constants without explicitly including the substrate into the supercells.
In order to speed-up the convergence, an initial value for the relaxed lattice constant in $z$-direction was set according to the formula $\frac{(x \cdot a_{\text{GaBi}} + (1-x) \cdot a_{\text{GaSb}})^3}{a_{\text{GaSb}}^2}$, assuming that the relaxation in $z$-direction will enhance the volume by the same factor as an interpolation between the lattice constants of the pure compounds in all directions would.
Using the LDA-1/2 potentials obtained for Sb and Bi, energy calculations were performed on the supercells. First, a self-consistent calculation with the regular grid of $k$-points described above was performed to obtain the charge densities and bandgap energies. 
Second, non self-consistent calculations using the charge densities were evaluated with finer $k$-point grids along a high-symmetry line in the first Brillouin zone from the $\Gamma$- to L-Point to obtain the bandstructure and Coulomb and dipole matrix elements. Spin-orbit coupling was included for all energy calculations.
\section{Results and Discussion}

In TABLE~\ref{tab:lattconst}, we summarize the lattice constants for the different III-V binaries as obtained from DFT calculations within the LDA. Comparing our values with those found in the literature \cite{Vurgaftman2001}, we obtain good agreement with the exception of GaP. Moreover, our computed lattice constants for the III-Bi binaries compare well to the values found in the literature \cite{Wang2002, Carrier2004, Ferhat2006}. It should be noted, that bulk InBi is reported to crystallize in the PbO structure \cite{Ferhat2006}, however, here we are interested in incorporating Bi into zinc-blende lattices and therefore, the lattice constant from the zinc-blende InBi structure is given in TABLE~\ref{tab:lattconst}.

\begin{table}
\caption{Lattice constants $a$ in \AA$\,$ calculated within the LDA for the zinc-blende structures of AlN, AlP, AlAs, AlSb, AlBi, GaN, GaP, GaAs, GaSb, GaBi, InN, InP, InAs, InSb and InBi. In addition to the LDA lattice constants, the deviations $\Delta_\text{exp}$ to the experimental lattice constants are listed.}\label{tab:lattconst}
\begin{ruledtabular}
\begin{tabular}{lccccc}
$X$ & N & P & As & Sb & Bi \\
\hline
$a($Al$X)$ & 4.3435 & 5.4337 & 5.6351 & 6.1155 & 6.2741 \\
$\Delta_\text{exp}^{\text{Al}X}$ & $-0.8\,\%$ & $-0.6\,\%$ & $-0.5\,\%$ &  $-0.3\,\%$ & $-$ \\[.7ex]
$a($Ga$X)$ & 4.4570 & 5.3208 & 5.6000 & 6.0529 & 6.2538 \\
$\Delta_\text{exp}^{\text{Ga}X}$ & $-1.0\,\%$ & $-2.4\,\%$ & $-0.9\,\%$ & $-0.7\,\%$ & $-$\\[.7ex] 
$a($In$X)$ & 4.9412 & 5.8097 & 6.0071 & 6.4304 & 6.6081 \\
$\Delta_\text{exp}^{\text{In}X}$ & $-0.8\,\%$ & $-1.0\,\%$ & $-0.9\,\%$ & $-0.8\,\%$ & $-$\\
\end{tabular}
\end{ruledtabular}
\end{table}

\subsection{Determination of the cutoff radii}

The cutoff radii for the anions of the zinc-blende structures for Al$X$, Ga$X$ and In$X$ for $X \in\,$\{N, P, As, Sb\} were found by adapting the procedure described in Ref.~\onlinecite{Ferreira2008}. Thereby, several pseudopotentials with different $r_\text{cut}$ for the anions were generated for each material system. The cutoff radii $r_\text{cut}$ were varied in steps of $0.1\,a_0$, where $a_0$ denotes the Bohr radius, and the direct bandgap at the $\Gamma$-point was extracted from the calculations. 
A parabolic function $E_g(r_\text{cut}) = m(r_\text{cut}-r_\text{cut}^\text{max})^2 + E_g^\text{max}$ was fit to the results of the calculations with the adjustable parameters $m$, $E_g^\text{max}$ and $r_\text{cut}^\text{max}$.
The maximum of this function is located at $r_\text{cut} = r_\text{cut}^\text{max}$, which is the optimized cutoff radius for the respective material.

\begin{figure}[htbp]
  \includegraphics[width=\columnwidth]{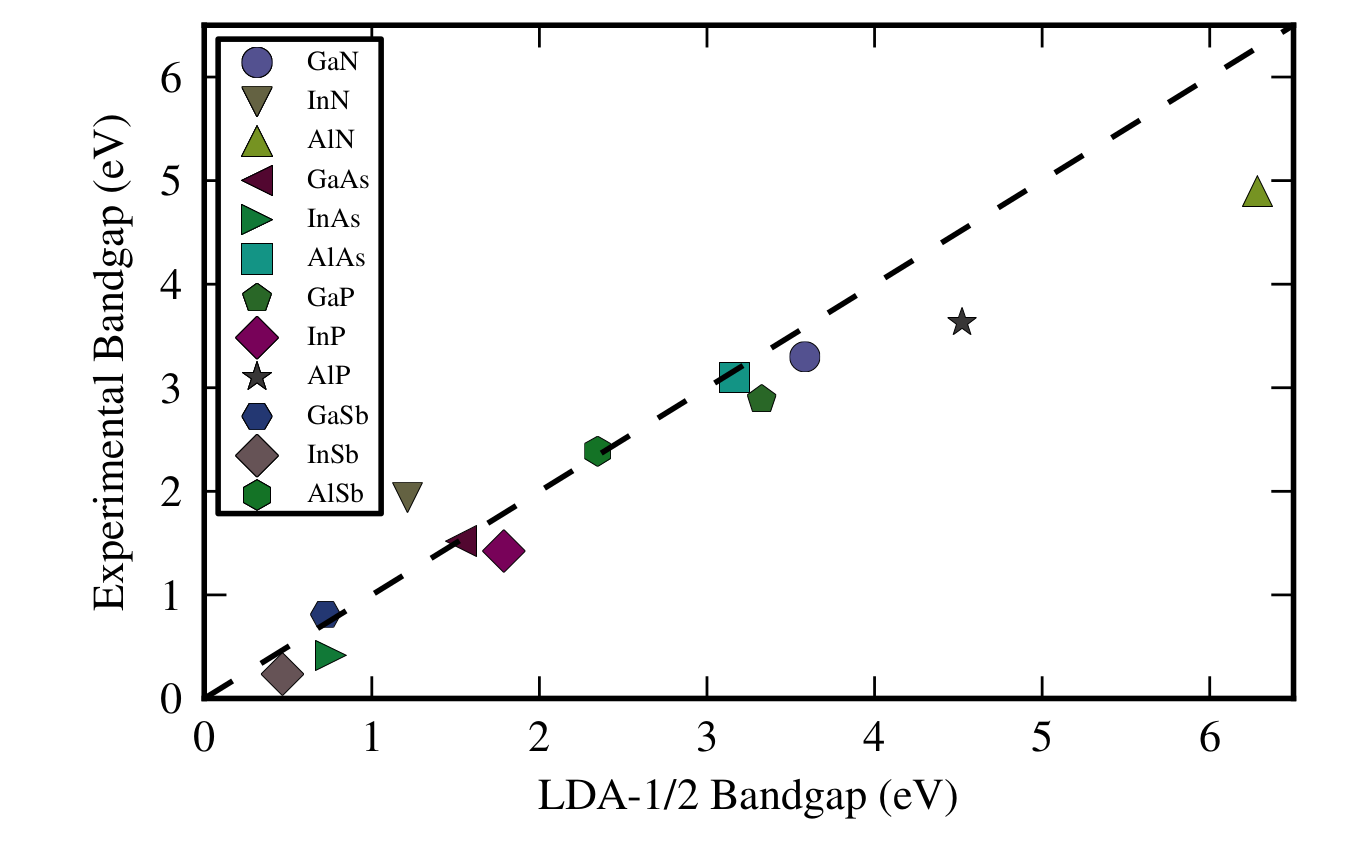}
  \caption{Experimental bandgaps for $T = 0\,$K taken from Ref.~\onlinecite{Vurgaftman2001} compared to the bandgaps as obtained from the LDA-1/2 method for several III-V semiconductors. Each material is represented by a data point with the experimental bandgap energy as $y$-value and the calculated bandgap energy as $x$-value, so that data points close to the straight line correspond to an almost perfect agreement for the bandgaps. Shown are the direct bandgaps at the $\Gamma$-point, even though some materials, i.e. AlN AlP, GaP, AlAs and AlSb are predicted to be indirect semiconductors.}\label{fig:bandgaps}
\end{figure}

In TABLE~\ref{tab:cutoffradii} the cutoff radii yielding the maximum bandgaps are shown together with the covalent radii for the anions. In comparison with Ref.~\onlinecite{Ferreira2008}, here the cutoff radii are smaller which is due to the use of $n=20$ for the spherical step function as compared to $n=8$ in Ref.~\onlinecite{Ferreira2008}. In order to obtain suitable cutoff radii for the Bi atoms, the cutoff radius as a function of the covalent radius for the group V atoms N, P, As and Sb is extrapolated. This yields the radii for Bi listed in TABLE~\ref{tab:cutoffradii}. \\
\begin{table}
\caption{Cutoff radii $r_\text{cut}$ as obtained for the zinc-blende structures of AlN, AlP, AlAs, AlSb, GaN, GaP, GaAs, GaSb, InN, InP, InAs, InSb together with the covalent radii for the five anions. The cutoff radii for the III-Bi materials were obtained using a linear extrapolation as described in the text. Here, $a_0$ denotes the Bohr radius.}\label{tab:cutoffradii}
\begin{ruledtabular}
\begin{tabular}{lccccc}
 & N & P & As & Sb & Bi \\
\hline
$r_\text{cov}/a_0$ & 1.342 & 2.022 & 2.249 & 2.627 & 2.797 \\
$r_\text{cut}$(Al)$/a_0$ & 2.573 & 3.329 & 3.403 & 3.749 & 3.934 \\
$r_\text{cut}$(Ga)$/a_0$ & 2.565 & 3.195 & 3.326 & 3.661 & 3.812 \\
$r_\text{cut}$(In)$/a_0$ & 2.737 & 3.351 & 3.454 & 3.865 & 3.986 \\
\end{tabular}
\end{ruledtabular}
\end{table}
Fig.~\ref{fig:bandgaps} compares the experimental values\cite{Vurgaftman2001} and those from the LDA-1/2 calculations for the direct bandgaps at the $\Gamma$-point for all investigated binary III-V semiconductors. 
Almost perfect agreement is found for GaSb, GaAs, AlSb and AlAs while all other bandgaps with the exception of InN are overestimated by the LDA-1/2 method. One reason for this overestimation is that LDA was used to calculate the lattice constants and LDA is known to yield smaller values than those reported in the experiment. Nonetheless, overall a strong improvement as compared to the LDA and PBE bandgaps is evident. \\
Applying the LDA-1/2 method to dilute bismides, the bandgap for GaAs$_{1-x}$Bi$_x$ was calculated for five 128 atom supercells with $x = 1.5625...7.8125\,\%$. The results are compared to bandgap values obtained with the TB09 functional in Ref.~\onlinecite{Bannow2017b}. It is found that the absolute values between both sets of calculations have a maximal difference of $\Delta E_g = 0.03\,$eV which is well within the accuracy of a specific configuration of Bi. 
Moreover, the LDA-1/2 bandgap decreases by $75\pm6\,$meV per 1$\,\%$ Bi which compares extremely well to the $72\pm4\,$meV per 1$\,\%$ Bi found in Ref.~\onlinecite{Bannow2017b}.
\subsection{Bandgap narrowing in Ga(SbBi)}
\begin{figure}[htbp]
    \includegraphics[width=\columnwidth]{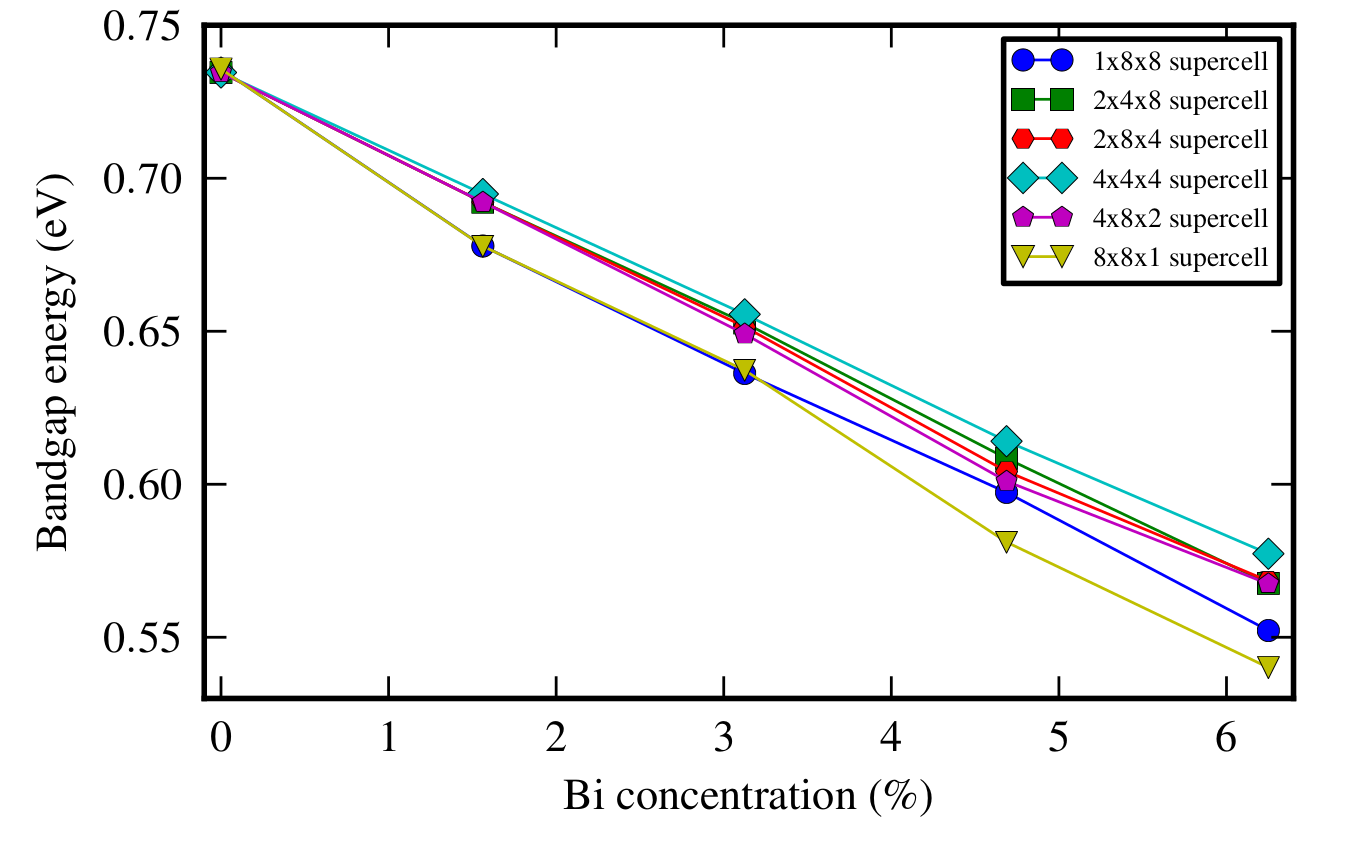}
    \caption{Theoretical bandgap energies from DFT calculations of GaSb$_{1-x}$Bi$_x$ supercells against Bi concentration $x$ using the LDA-1/2 potentials. The supercells consist of 64 primitive cells, which are arranged in different cuboidal geometries. The lines are guides to the eye.}\label{gasbbi_bandgaps}
\end{figure}

\begin{figure}[htbp]
    \includegraphics[width=\columnwidth]{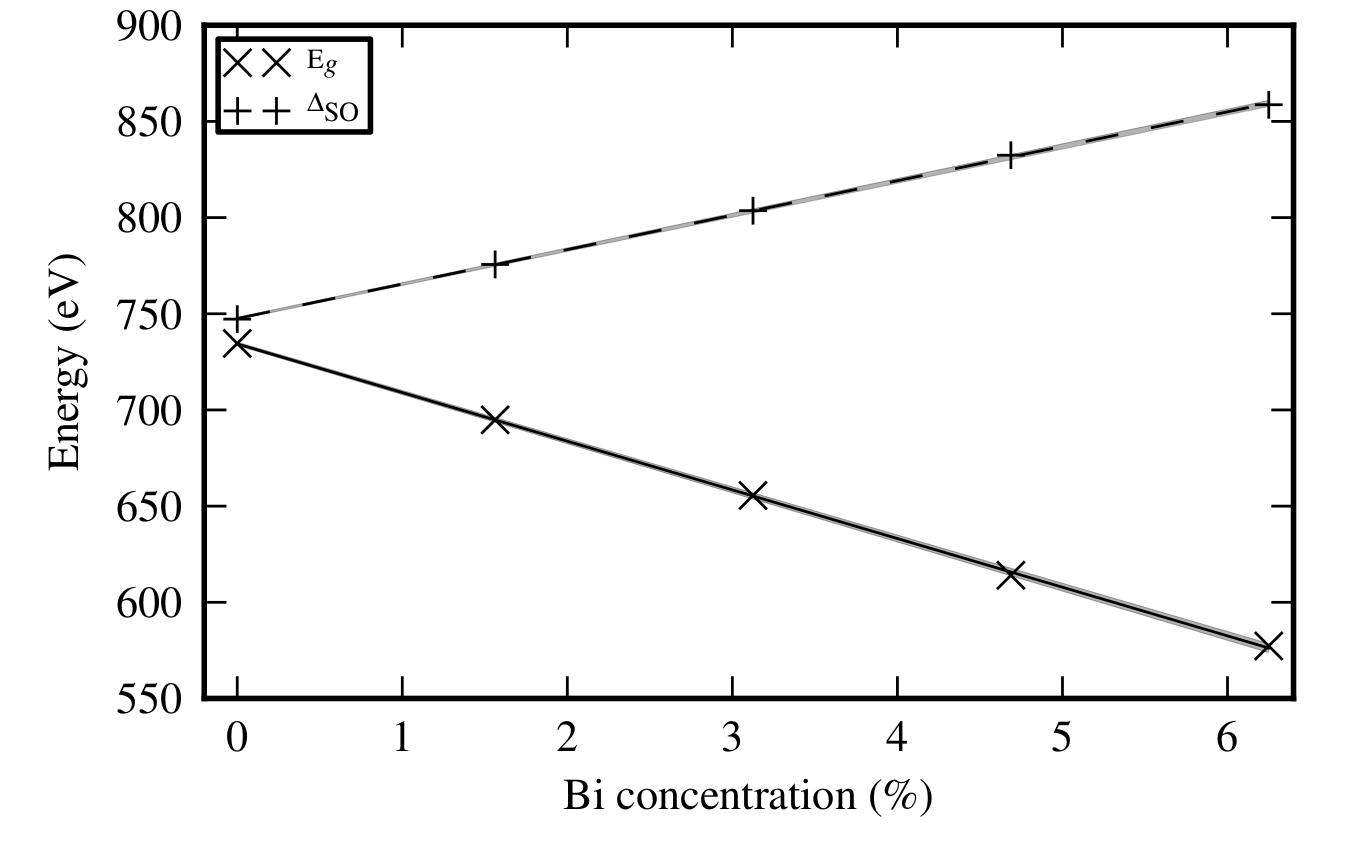}
    \caption{Bandgap and spin-orbit splitting energy ($\Delta_\text{SO}$) of the GaSb$_{1-x}$Bi$_x$ supercells consisting of 4x4x4 primitive cells over Bi concentration $x$ obtained from DFT calculations using the LDA-1/2 method. The lines represent linear fits with the shaded areas indicating the standard deviations.}
    \label{gasbbi_eg_so}
\end{figure}

Using the obtained LDA-1/2 potentials for GaSb and GaBi, supercell calculations on GaSb$_{1-x}$Bi$_x$ were performed as described in Sec. \ref{sup}.
A comparison of the bandgap energies with increasing Bi content for the different supercell geometries is shown in Fig.~\ref{gasbbi_bandgaps}.
An almost linear decrease of the bandgap can be seen for all supercell geometries. 
For all concentrations, the bandgap energy of the isotropic 4x4x4 supercell is the highest. 
The supercells with 2 primitve cells in one direction, 2x4x8 and its permutations, behave similarly and have slightly lower bandgap energies than the isotropic supercell.

The strongest, not linear decrease in bandgap energy is found for the supercells that have only one primitive cell in one of the spatial directions. However, 
this behavior is an artifact resulting from to the periodic boundary conditions of the DFT calculations which impose an artificial ordering by creating an infinite line of neighboring Bi atoms along the direction in which only one primitive cell is used.
This induces spurious interactions between the Bi atoms and their images, which causes an effect on the bandgap exceeding that in a real material where the Bi atoms are spaced more evenly throughout the crystal. 

Therefore, we conclude that the isotropic supercells with 4x4x4 primitive cells yield the best results by minimizing spurious interactions. 
The bandgap energies and split-off energies of these supercells are shown in Fig.~\ref{gasbbi_eg_so}.
A linear fit yields a bandgap reduction of $(25.3 \pm 0.2)$ meV/\%Bi and an increase of the spin-orbit splitting energy of $(17.9 \pm 0.2)$ meV/\%Bi. 
The spin-orbit splitting is in good agreement with 18meV/\%Bi found in a theoretical study\cite{Polak2015} and the bandgap narrowing is comparable to 30meV/\%Bi observed in an experimental study\cite{Kopaczek2013}.
For the following analysis of the bandstrucutre and optical properties, only the isotropic 4x4x4 supercells are considered. 

\subsection{Effective Bandstructure of Ga(SbBi)}
\begin{figure}[htbp]
    \includegraphics[width=\columnwidth]{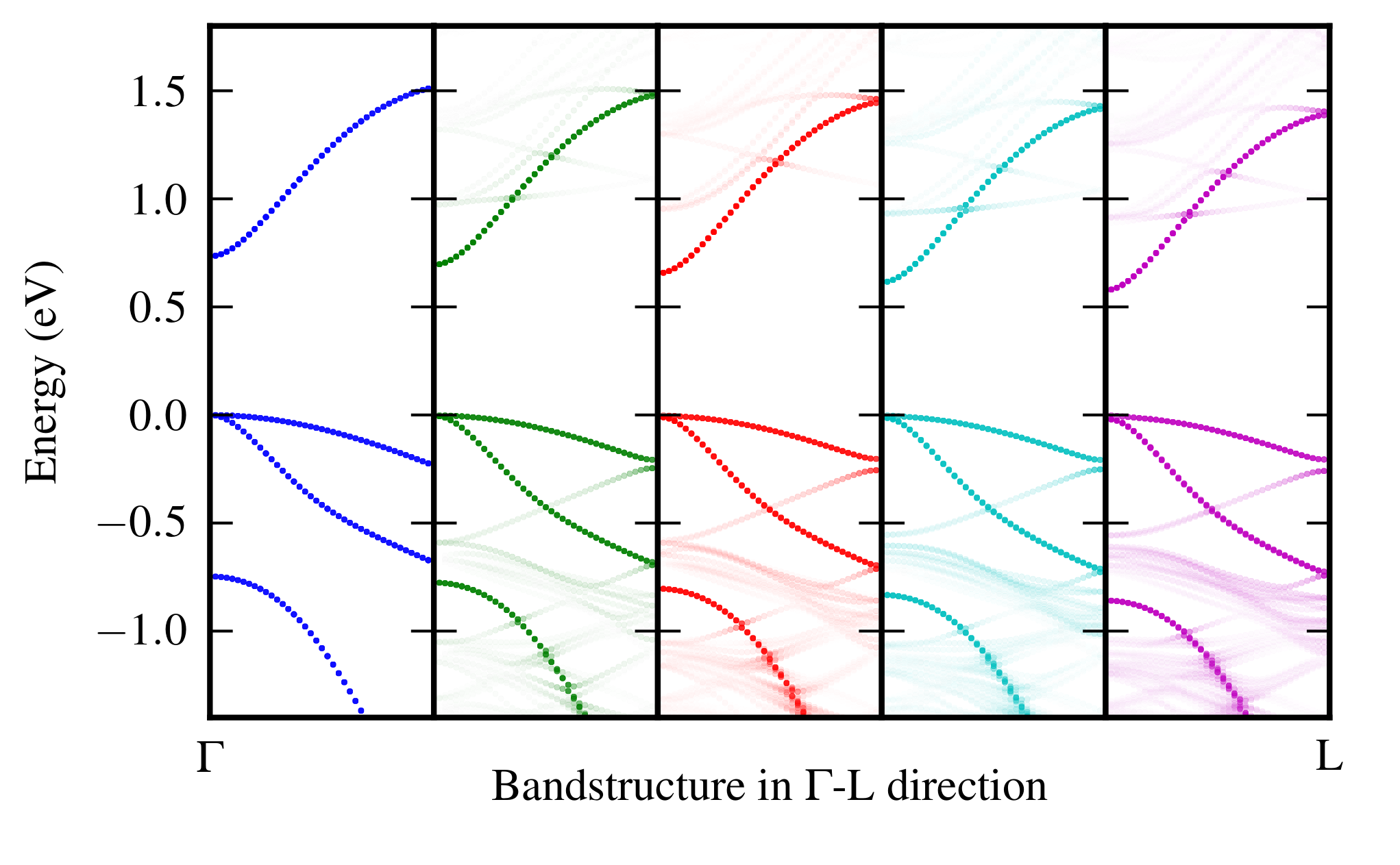}
    \caption{Bandstructures from $\Gamma$- to L-point of GaSb$_{1-x}$Bi$_x$ supercells for different Bi concentrations, from left to right $x=0\%, 1.56\%, 3.13\%, 4.69\%, 6.25\%$. The electronic structure has been calculated using DFT with the LDA-1/2 method and the supercell bandstructure was unfolded to obtain the effective primitive cell bandstructure. The opacity of the data points indicates their spectral weight.}
    \label{gasbbi_ebs}
\end{figure}

The repetition of primitive cells in real space leads to an backfolding of states in $k$-space towards the zone center. 
Therefore, in order to compare the bandstructure of a supercell with that of the primitve cell, it is convenient to perform an unfolding of the bands to obtain the effective bandstruture\cite{popescu_ebs}.
The spectral weight of the unfolded states represents how much of the Bloch character of the state is preserved in the supercell.
The unfolded bandstructures of the isotropic supercells for all Bi concentrations (increasing from left to right) are shown in Fig.~\ref{gasbbi_ebs}, where the opacity of the scatter points was set according to the spectral weight of the respective states.
Therefore, states with lower spectral weight are less visible than those with higher spectral weight.
The energies are given relative to the valence band edge, so that the valence band maximum is always at 0 eV.

In the pure bandstructure, the light and heavy hole band, the split-off band, and the conduction band can be seen. 
These bands remain the ones with the highest spectral weight for all concentrations and can be clearly identified.
However, we notice a shift of the conduction and the split-off band to lower energies with increasing Bi concentration.
At the same time, more and more additional bands with low spectral weight emerge.
In a region between the valence bands and the split-off band, a lot of defect bands are very close to one another.
Additionally, the spectral weight of the defect bands increases for higher Bi concentrations.

At two points, the conduction band is crossed by defect bands with low spectral weight. 
Near the crossing points, the spectral weight of the defect bands increases.
The spectral weight of some defect bands also increases towards the zone edges. 
A slight splitting of the heavy- and light-hole band can be seen for higher concentrations. 
This splitting can be quantified approximately using an exponential fit, yielding
\begin{align}
 E_{hh-lh} (x) = (3.5 \pm 0.5) \, \text{meV} \cdot \text{exp} \left( (27 \pm 3) x \right)  \,
\end{align}
for ${\rm{GaSb}}_{1-x}{\rm{Bi}}_{x}$.
\subsection{Optical Properties of Ga(SbBi)}
\begin{figure}[htbp]
    \includegraphics[width=\columnwidth]{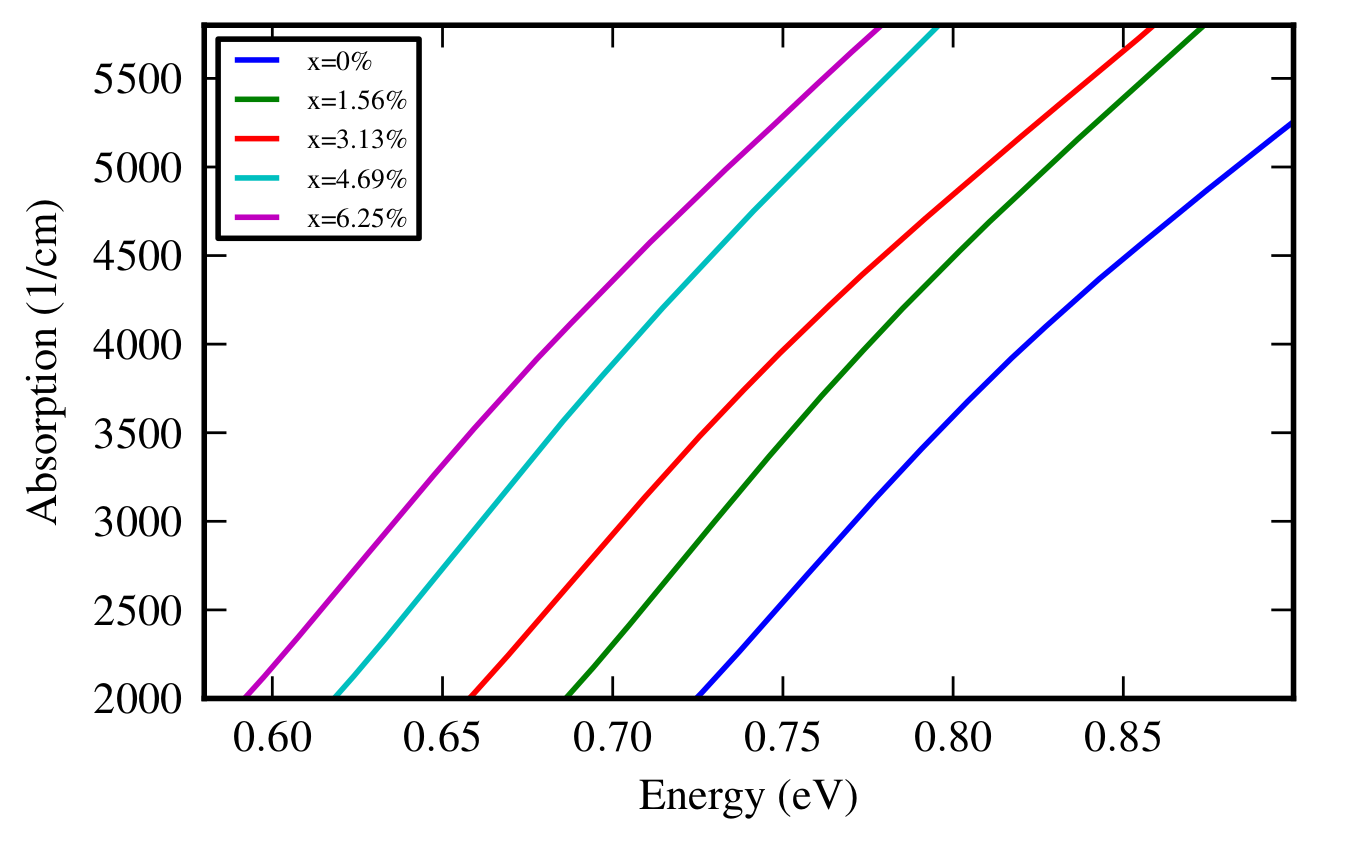}
    \caption{Absorption spectra of ${\rm{GaSb}}_{1-x}{\rm{Bi}}_{x}$ supercells for different Bi concentrations as a function of photon energy at 300K with carrier density $N_{3D}=5 \cdot 10^{16}$cm$^{-3}$. Dipole and matrix elements from DFT calculations using the LDA-1/2 method were used as input for a microscopic theory.}
    \label{gasbbi_abs}
\end{figure}
The DFT wavefunctions are used to compute the Coulomb and dipole matrix elements needed for our microscopic Semiconductor Bloch Equation approach to calculate the intrinsic optical properties~\cite{Haug2009}.
The details of the calculation can be found in Refs.~\onlinecite{lindberg1988,girndt1997}.
In Fig.~\ref{gasbbi_abs}, the resulting absorption spectra for the different Bi concentrations are shown.
For the calculations, 20 conduction bands and 30 valence bands were included.
The temperature was set to 300K and the carrier density was set to $N_{3D}=5 \cdot 10^{16}$cm$^{-3}$.
It should be noted that the DFT bandstructure was used, altough it was calculated at 0K temperature.
As expected, an increase in the absorption coefficient can be seen at the bandgap energy. 
For higher Bi concentrations, this increase shifts to lower energies, in agreement with the decreasing bandgap.

\begin{figure}[htbp]
    \includegraphics[width=\columnwidth]{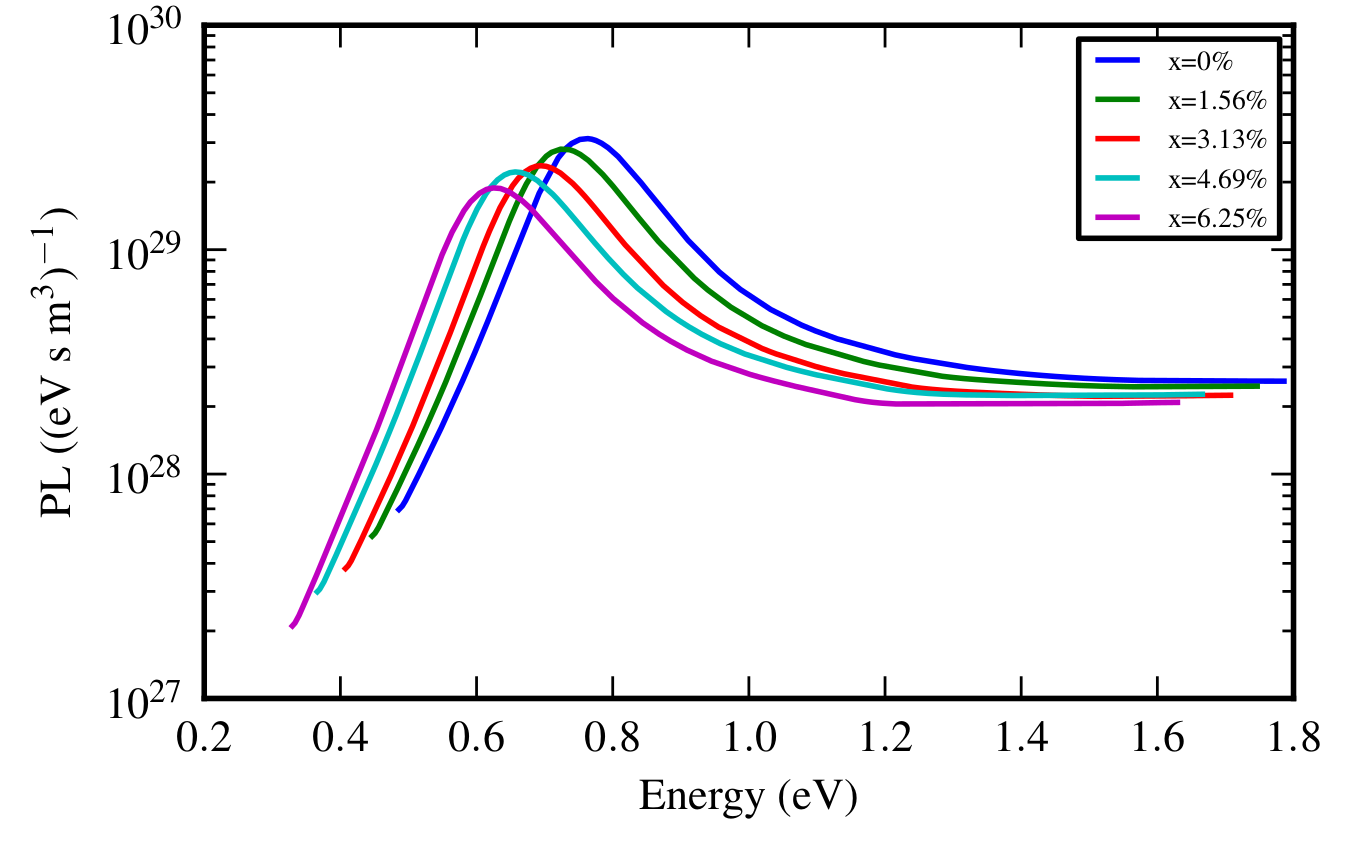}
    \caption{Photoluminescence spectra of ${\rm{GaSb}}_{1-x}{\rm{Bi}}_{x}$ supercells for different Bi concentrations as a function of photon energy at 300K with carrier density $N_{3D}=5 \cdot 10^{16}$cm$^{-3}$. Dipole and matrix elements from DFT calculations using the LDA-1/2 method were used as input for a microscopic theory.}
    \label{gasbbi_pl}
\end{figure}
In order to calculate the emission properties, the microscopic approach is used to solve the Semiconductor Luminescence Equations~\cite{kira2006,kira2006pqe}.
Photoluminescence spectra at 300K for the supercells have been calculated and are shown in Fig.\ref{gasbbi_pl}.
Again, 20 conduction bands and 30 valence bands were included in the calculation.
An intensity peak for each concentration occurs at the respective bandgap energy.
Therefore, a shift to lower energies can be seen for increasing Bi concentration.

\subsection{Auger losses of Ga(SbBi)}
\begin{figure}[htbp]
    \includegraphics[width=\columnwidth]{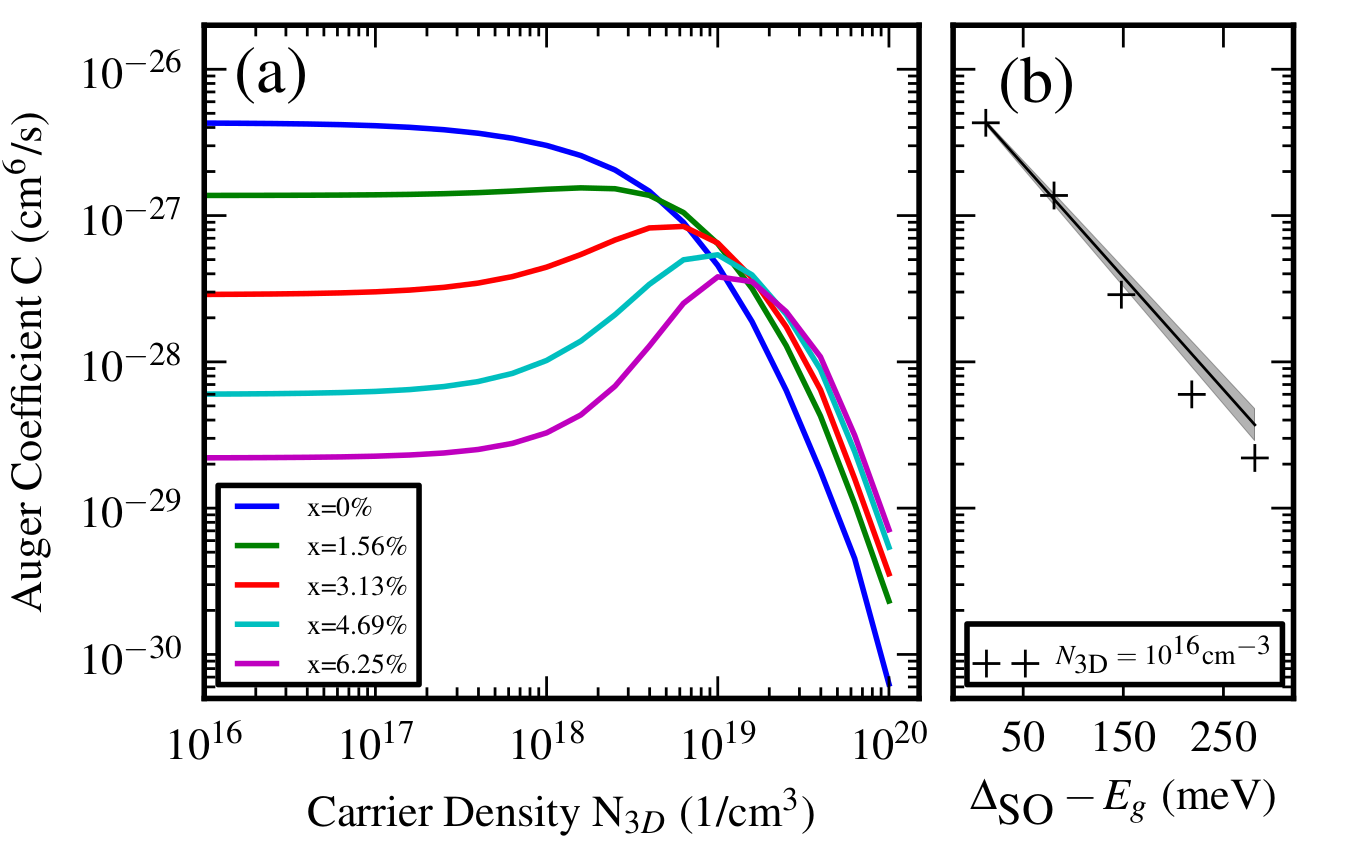}
    \caption{(a) Auger loss coefficients of ${\rm{GaSb}}_{1-x}{\rm{Bi}}_{x}$ supercells for different Bi concentrations as a function of carrier density. Dipole and matrix elements from DFT calculations using the LDA-1/2 method were used as input for a microscopic theory to calculate the Auger losses. (b) Auger loss coefficients as a function of the difference between spin-orbit splitting and bandgap energy. The line represents an exponential fit with the shaded area indicating the standard deviation.}
    \label{gasbbi_auger}
\end{figure}
The losses due to Auger recombination at 300K have been calculated within the microscopic approach as described in Ref.~\onlinecite{hader2005jqe} by solving Eqs. $(9-10)$ therein and using Eq. $(11)$ to obtain the Auger loss coefficients. 
The energy region from 1.1 eV below the valence band maximum up to 1.7 eV above the valence band maximum was used for the calculations.
The results for different carrier densities are shown in Fig.~\ref{gasbbi_auger}.

For low densities up to $N_{3D}=10^{17}$cm$^{-3}$, we see that the Auger coefficient is virtually constant for all Bi concentrations considered.
The coefficient is highest for pure GaSb with $4.3 \cdot 10^{-27} \rm{cm}^6/\rm{s}$, which compares well to the experimental value of $(12\pm4) \cdot 10^{-27} \rm{cm}^6/\rm{s}$ found in Ref. \onlinecite{marchetti2002}, and decreases for higher Bi content.
Since for pure GaSb the bandgap and spin-orbit splitting energy are very close to each other (see Fig. \ref{gasbbi_eg_so}), the losses are high.
With increasing Bi content, the bandgap decreases, while the spin-orbit splitting increases, so that the loss channel is weaker and the Auger coefficient decreases.
The dependence of the Auger coefficient on the difference between bandgap and spin-orbit splitting energy in this low density region is shown in part (b) of Fig.~\ref{gasbbi_auger}, exhibiting a near exponential decrease, which can be approximated by
\begin{align}
 C  = (5.4 \pm 0.1) \cdot 10^{-27} \, \text{cm}^6\text{s}^{-1} \cdot \text{exp} \left( (17 \pm 1) \frac{\Delta_{\textrm{SO}}-E_g}{\text{eV}} \right)  \, .
\end{align}

For higher densities, from $N_{3D}=10^{18}$cm$^{-3}$ to $N_{3D}=10^{19}$cm$^{-3}$, more states are filled with electrons and more transitions become possible, so that the Auger coefficient rises.
The rise only occurs for the GaSbBi supercells, not for pure GaSb, and is more prominent for higher Bi concentrations.
However, at the same time, from $N_{3D}=10^{18}$cm$^{-3}$ on, the filling of the phase-space leads to a general decrease of the Auger coefficient for all Bi concentrations\cite{hader2005}.
The rise in the loss coefficient for the GaSbBi supercells is stronger for densities up to $N_{3D}=10^{19}$cm$^{-3}$, so that the Auger coefficient reaches a maximum.
For densities higher than $N_{3D}=10^{19}$cm$^{-3}$, the phase-space filling effect dominates and the Auger coefficients for all supercells decrease exponentially.
For a given density above $N_{3D}=10^{19}$cm$^{-3}$, pure GaSb has the lowest loss coefficient, and higher Bi content leads to higher loss coefficients.
It should be noted that the Auger coefficients have been calculated using only the Coulomb matrix elements of the $\Gamma$-L direction and results may differ in other directions.

\section{Conclusion}
The LDA-1/2 method for DFT calculations requires less memory than other exchange-correlation functionals while yielding good results for most semiconductor compounds.
Up to date, the LDA-1/2 method could not be used for dilute bismides, because the construction of the LDA-1/2 potential involves maximizing the bandgap of the compound.
In this work, a new method of finding the cut-off radius for the construction of LDA-1/2 potentials by extrapolation has been introduced. 
It was used to construct LDA-1/2 potentials for the semi-metal Bi in semiconductor compounds AlBi, GaBi and InBi, allowing for DFT calculations on dilute bismides with significantly reduced computational cost. In consequence, larger supercells, which are needed for low Bi concentrations, and dense $k$-point grids can be calculated.
The usefulness of the approach for dilute bismides was demonstrated by calculating the electronic and optical properties of Ga(SbBi) for different Bi concentrations by using the results from the DFT calculations with the LDA-1/2 method as input to a microscopic theory.

\begin{acknowledgments}
The Marburg work was supported by the DFG in the framework of the Research Training Group ``Functionalization of Semiconductors'' (GRK~1782). The authors thank the HRZ Marburg and CSC-Goethe-HLR Frankfurt for computational resources.
The Tucson work was supported by the Air Force Office of Scientific Research under award number FA9550-17-1-0246.
\end{acknowledgments} 

\section*{Data Availability}
The data that support the findings of this study are available from the corresponding author upon reasonable request.

\bibliography{literature}
\end{document}